# Relaxed or Tense? Mutual Biosignal Transmission with Heartbeat Vibrations during Online Gameplay

Yoshiki Mori, Sohta Takada, Masayuki Kajiura, Modar Hassan, and Taku Hachisu

*University of Tsukuba, Ibaraki, Japan*

(Email: {mori, takada, kajiura}@ah.iit.tsukuba.ac.jp, modar@ai.iit.tsukuba.ac.jp, hachisu@ah.iit.tsukuba.ac.jp)

**Abstract ---** Esports offers a platform for players to engage in competitive and cooperative gaming with others remotely via the Internet. Despite these opportunities for social interaction, many players may still experience loneliness while playing online games. This study aims to enhance the social presence of partner players during online gameplay. The demonstration system, designed for 1-on-1 online competitive games, mutually transmits the partner's biosignals, through heartbeat-like vibrotactile stimuli. The system generates vibrotactile signals that represent two-dimensional emotions, arousal, and valence, based on biosignals such as heart rate and electrodermal activity.

## 1 INTRODUCTION

Esports provides a platform for players to engage in competitive and cooperative gaming with individuals remotely via the Internet. This ensures equal playing conditions regardless of age, gender, nationality, or disability, fostering community building and social interactions among a diverse population.

Despite these opportunities for social interaction, many players may experience loneliness while playing online games. This loneliness may stem from the lack of robust social ties and direct physical interactions, as establishing close relationships is challenging when partners remain anonymous or the connections are temporary. Video calls during gameplay have been shown to facilitate interactions, but comprehensive engagement is limited because the visual and auditory senses are focused on the game. Additionally, poor video and audio quality of the calls due to the network traffic further impedes effective interactions.

This study aims to enhance the social presence of partner players during online gameplay. The demonstration system, designed for 1-on-1 online competitive games, mutually transmits the partner's biosignals through heartbeat-like vibrotactile stimuli (Fig.1). The system generates vibrotactile signals representing two-dimensional emotions: arousal (high arousal – sleepiness) and valence (pleasant feelings – unpleasant feelings) [1], based on biosignals such as heart rate (HR) and electrodermal activity (EDA). The system builds on [2] and has been revised to implement this heartbeat-like vibrotactile feedback function into the bracelet devices.

## 2 SYSTEM

We define three design requirements for the system to demonstrate its potential: 1) real-time biosignal measurement; 2) real-time biosignal transmission to a remote player through heartbeat-like vibrations; and 3) compatibility with daily gameplay. The heartbeat-like vibrations are intended to evoke emotional experiences and enhance social presence. For instance, higher heart rates are associated with increased arousal levels [3]. We enhance the emotional expression of these vibrations to represent valence by modifying the frequency of audio signals.

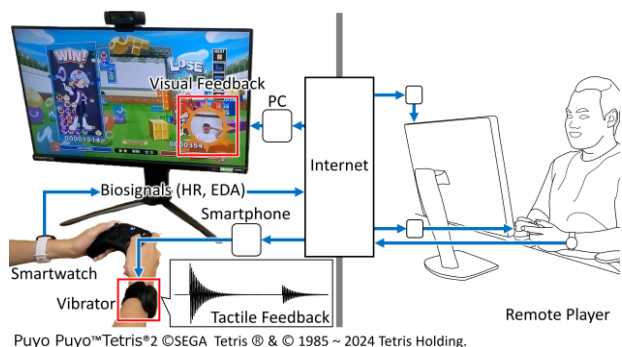

Puyo Puyo™ Tetris®2 ©SEGA Tetris ® & © 1985 ~ 2024 Tetris Holding.

Fig.1 A system for real-time mutual transmission of players' biosignals with heartbeat-like vibrations during online gameplay

As shown in Fig.1, the system comprises a pair of smartwatches, smartphones, bracelet devices, monitors, cameras, controllers, and host computers. The smartwatch (Google LLC, Google Pixel Watch 2) measures HR and EDA values and uploads them to a real-time database on the Internet (Google LLC, Firebase Realtime Database). The smartphone (Google LLC, Google Pixel 5a (5G)) downloads the remote player's HR and EDA values from the database. Based on these values, it generates audio signals $V$ at time $t$:

with
$$V(t') = V_1(t') + V_2(t'),$$
$$V_1(t') = A \exp(-Bt') \sin(2\pi f t'),$$
$$V_2(t') = \begin{cases} 0 & \text{if } 0 < t' < \varphi \\ CV_1(t' - \varphi) & \text{otherwise,} \end{cases} \quad (1)$$
$$t' = t - \frac{60}{H} \text{floor}\left(\frac{t}{60/H}\right)$$

where $V_1$ and $V_2$ are the first and second heart sounds (S1 and S2), respectively. These shape heartbeat-like signals consist of two impulse responses caused by the opening and closing of heart valves [4][5]. The parameters $A$, $B$, $C$, and $\varphi$ are the initial amplitude, decay rate, and amplitude ratio of S1 and S2, and onset asynchrony between S1 and S2, respectively, and are set to constant values empirically determined. The parameters $H$ and $f$ are the HR and frequency, respectively. Preliminary tests showed that increases in $H$ and $f$ evoke higher arousal and lower valence, respectively. We set $H$ and $f$ to five discrete levels (60, 75, 90, 105, 120 BPM and 100, 150, 200, 250, 300 Hz) based on measured HR and EDA, respectively. The system requires a calibration process to measure biosignals during rest and stress, but this demonstration omits this process due to time constraints.

The bracelet-type device comprises a case with a vibrator (Acouve Laboratory, Vp210), a wristband, and an audio amplifier (North Flat Japan, FX202A/FX-36APRO). The audio amplifier amplifies the audio signals from the smartphone and drives the vibrator.

In addition to the game application, the host computer runs software that visualizes the five levels of the remote player's HR and EDA values around the remote player's face video, displaying this in the bottom right corner of the game screen. This visualization allows for the comparison with the vibrotactile stimuli.

## 3 Demonstration

Participants experience their partner's biosignals through heartbeat-like vibrations during gameplay.

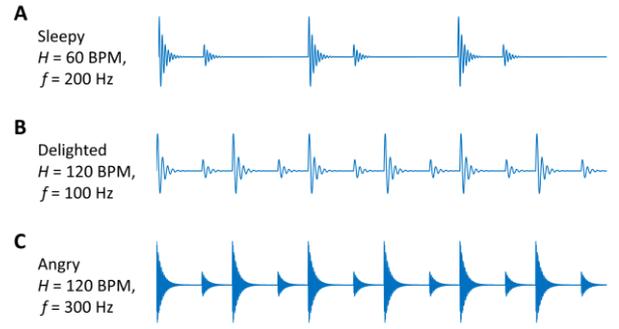

Fig.2 Examples of vibrotactile signals: A) sleepy; B) delighted; C) angry

Participants are first instructed to be seated in front of the monitor and wear a smartwatch on their left wrist, and a bracelet device on their right wrist. Before the gameplay begins, a demonstrator explains the vibrotactile signals by playing some examples, illustrating the relation between the signals and the emotions they represent. Participants then start playing the game using a controller (Fig.1). Figure 2 shows examples of vibrotactile signals. When the emotion is sleepy, the signal appears as shown in Fig.2A. For a delighted emotion (high arousal and pleasure), $H$ increases and the $f$ decreases, as depicted in Fig.2B. For an angry emotion (high arousal and displeasure), both $H$ and $f$ increase, as shown in Fig.2C. Each session lasts 3 minutes.

Acknowledgement

Funding: This work was supported by the New Energy and Industrial Technology Development Organization (NEDO) (JPNP21004).